# Time Synchronization over a Free-Space Optical Communication Channel

ISAAC KHADER,[1*] HUGO BERGERON,[3] LAURA C. SINCLAIR,[1] WILLIAM C. SWANN,[1] NATHAN R. NEWBURY[1] AND JEAN-DANIEL DESCHÊNES[2*]

[1]National Institute of Standards and Technology, 325 Broadway, Boulder, CO, 80305, USA
[2]Octosig Consulting Inc., 208-3035 rue des Chatelets, Quebec, QC, G1V 3Y7, Canada
[3]Université Laval, 2325 Rue de l'Université, Québec, QC, G1V 0A6, Canada

*Corresponding authors: isaac.khader@nist.gov, octosigconsulting@gmail.com

Free space optical (FSO) communication channels are typically used to transmit high-speed data between sites over the air. Here we repurpose an FSO digital communication system and use it directly for two-way time transfer. We demonstrate real-time synchronization between two sites over a turbulent air path of 4 km using binary phase modulated CW laser light. Under synchronization, the two sites exhibit a fractional frequency deviation below $10^{-15}$ and a time deviation below 1 picosecond at averaging times of seconds to hours. Over an 8-hour period the peak to peak wander is 16 ps.

OCIS codes: (060.2605) Free-space optical communications; (060.1155) All-optical networks; (120.3940) Metrology;

## 1. Introduction

Free space optical (FSO) communications systems are an appealing alternative to conventional free space RF communications due to the higher available data rates for applications such as cellular back-haul, fiber backup during disaster recovery, metropolitan area networks, secure communications, and satellite links [1]. Here we explore the use of a digital FSO communication system for a different application, namely to support optical clock networks via time transfer. Many applications for optical clock networks are being studied including phased array telescopes, geodesy, and precise navigation and timing [2].

In the past, we have demonstrated optical two-way time-frequency transfer (OTWTFT) by the two-way transmission of coherent frequency comb pulse trains, combined with linear optical sampling. This comb-based OTWTFT has shown femtosecond level synchronization over free space and is therefore capable of supporting state-of-the-art optical clocks [3]. However, this system is complicated, requiring three frequency combs for a point-to-point link and the high-fidelity transmission of analog frequency comb pulses. Moreover, this level of timing precision is not always required. In this work, we present a time transfer method over a free space link that uses only a digital optical communication channel. This approach still exploits two-way time transfer (and therefore the reciprocity of the atmosphere) but requires no additional hardware beyond the FSO communication link. The FSO link is based on a half-duplex architecture, binary phase shift keying, and coherent heterodyne detection, but similar results should be achievable with any single-mode spatial communication link.

Until recently, free space time transfer has been dominated by microwave techniques. Microwave satellite methods achieve synchronization to the nanosecond level using spread-spectrum pseudorandom modulation on an RF carrier [4]. This performance is comparable to GPS common-view time transfer techniques [5]. GPS carrier phase methods have achieved performance at the tens of picosecond level [6].

Optical-domain time transfer techniques should be able to achieve significant performance improvements over these microwave methods. Indeed, optical time-frequency transfer over long-distance optical fiber-based links is now well-established [7,8]. Fiber links allow for long-distance picosecond and femtosecond-level time transfer [9,10] and $10^{-18}$ level frequency transfer [11,12]. However, these fiber-based approaches require dedicated, fixed fiber links. Thus, the development of free space alternatives is desirable.

Several different approaches have been explored for optical time-frequency transfer over free-space. The aforementioned comb-based OTWTFT has been shown to synchronize clock times to within femtoseconds over 12 km of air [13] and in the presence of Doppler shifts, but at the cost of system complexity. Recent one-way phase compensation methods have reached picosecond timing wander over 100 m distances [14]. Vastly longer distances have been demonstrated with satellite-based FSO time transfer methods. Sub-nanosecond timing has been achieved by bouncing optical pulses off of nanosatellites [15]. The pulse-based T2L2 and direct intensity-detection-based ELT methods have achieved 10 ps timing wander over a day [16,17]. While these techniques have been proven to work with good performance, here we explore an alternative spread-spectrum method.

The intent is to provide straightforward all-optical free space time transfer and synchronization at the picosecond level that relies on existing optical communication system designs. In many ways, the approach is analogous to microwave satellite time transfer [4] and more recent work over fiber-optic telecommunication networks [18]. The system uses phase-modulated, large time-bandwidth product waveforms and coherent detection over a reciprocal single-mode link. Compared to comb-based OTWTFT, this approach minimizes hardware overhead in systems with many stations and connections. Additionally, the phase-modulated CW laser light used here is simpler to amplify compared to comb pulses, which is important for long distance applications such as ground to satellite or intra-satellite communications systems [19,20]. With this system, we show it is possible to actively synchronize a remote clock to a master clock with a time deviation (TDEV) below 1 picosecond from seconds to hours of averaging time and a peak to peak wander of 16 ps over 8 hours.

Fig 1a shows the general layout of the synchronization system, consisting of a master and remote site. The clock at each site consists in this case of a cavity stabilized laser locked to a frequency comb, but could be a simple RF oscillator as well [21]. An optical coherent communication link connects the sites and is used both for OTWTFT and for data transfer. At the remote site, the OTWTFT timing data is used to calculate the clock offset in real-time, which is then input into a phase-locked loop to actively synchronize the remote clock to the master clock. Our frequency comb-based OTWTFT of Ref. [3] (accurate to a few femtoseconds) runs in parallel to verify the time synchronization achieved here with only the coherent communication based system.

## 2. Experimental Setup

We establish a digital optical coherent communication link between the sites based on binary phase-shift keyed (BPSK) modulation [22] of cw laser light. This digital communication channel is exploited for both timestamp generation and data transmission. The timestamps are transmitted via an extended waveform, comprising a pseudorandom binary sequence (PRBS) and local clock counter. The system runs in half-duplex mode, such that data can be transferred in both directions but not simultaneously. At the remote site, the local laser is loosely frequency locked to the incoming light so that the heterodyne signal is offset by about 250 MHz, falling in the detector bandwidth. To shift the modulation away from the cw carrier (and associated phase noise), Manchester encoding is used.

For OTWTFT, the system measures timestamps for four events occurring either at the master or the remote site. An example of this exchange is shown in Fig. 1b. We first measure the transmit time, $T_{AA}$, of the PRBS waveform from the master site according to the local master clock. After a delay due to the link distance, we record the arrival time of this extended waveform, $T_{BA}$, at the remote site according to the local remote clock. The same procedure is done for a waveform sent from the remote site, yielding two other event timestamps: $T_{BB}$ and $T_{AB}$, recorded according the remote and master clocks respectively. If this procedure is accomplished quickly enough (here in 400 microseconds), the transmitted waveforms both experience the same free-space turbulence, and therefore have equal link delays, due to the reciprocity of the single-mode link [23]. The communication link transfers the measured timestamps data from the master site to the remote site. Following the general prescription for two-way time transfer, the clock time difference is computed at the remote site as the following linear combination:

$$\Delta T_{AB} = \frac{T_{BA} - T_{AA}}{2} - \frac{T_{AB} - T_{BB}}{2} + \Delta T_{cal}. \quad (1)$$

Note that $\Delta T_{cal}$, includes fixed differential delays (due to fiber paths as well as photodetector and RF components) in the transceivers.

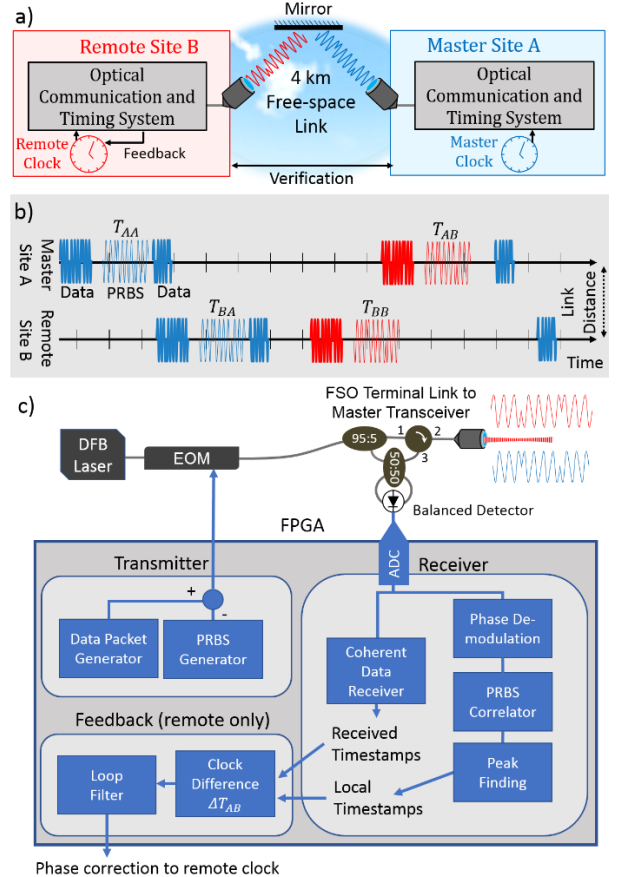

Fig. 1: a) High-level schematic of the experimental setup. A coherent communication channel is established over 4 km of air between the two sites. This two-way signal is used to synchronize the two clocks within picoseconds. b) Diagram of the full pseudorandom binary sequence (PRBS) and data exchange on the half-duplex link. For each PRBS, two timestamps are determined, one at each site. The signals launched from the Master site are shown in blue, while the ones launched from the Remote site are shown in red. Data packets are time-multiplexed across the same link and are shown in bold. c) Transceiver design for the remote site. The binary modulation is written onto the phase of the 1546-nm DFB laser through small angle phase modulation by an EOM. FPGA: Field Programmable Gate Array, FSO terminal: Free Space Optical terminal, EOM: Electro-Optic phase Modulator, DFB: Distributed Feedback.

Fig. 1c shows the details of the transceivers used for the timing exchange. A 100 mW distributed feedback (DFB) communications laser at 1546 nm provides the optical carrier. An electro-optic phase modulator (EOM) encodes the RF waveform on the carrier. The laser light is split by a 95:5 coupler. The 95% signal is launched while the 5% tap serves as a local oscillator for the received light. The heterodyne signal is detected using a balanced detector. This heterodyne signal is locked to ~250 MHz through feedback to the DFB laser temperature at the remote site. The entire system uses standard telecommunication components.

At each site, a Field Programmable Gate Array (FPGA) is used for packet data communication and timestamp generation. A general purpose digital IO pin of the FPGA drives the EOM to apply the binary phase shift modulation to the carrier. The received heterodyne signal is sampled by an ADC at 200 MS/s and processed on the FPGA.

To recover the serialized data from the data packet generator, the coherent data receiver module in the FPGA uses several standard techniques from optical communications [22]. First, we demodulate the phase of the received heterodyne beat at ~250 MHz. The data clock is recovered first using a short training sequence at the start of each data packet, then the word boundaries are recovered using a start word, finally the data packets contents are recovered. The data integrity is verified using a cyclic-redundancy check (CRC). No forward-error correction (FEC) was implemented as the synchronization system can tolerate dropped packets.

To modify the communications system for time transfer, a PRBS generation system was added on the same channel in a time-multiplexed fashion. While the data packets themselves could theoretically be used for timing purposes, a known PRBS is used as the modulation type to ensure large ambiguity range and narrow pulse correlation with a fixed template. The full sequence is 1024-chips with 100 ns chip duration (10 MHz chip rate). Each PRBS is transmitted along with a local 64-bit clock counter at the transmission time, yielding for all practical purposes an infinite ambiguity range of 5 ns x $2^{64}$ (equivalently ~3 x $10^{19}$m).

The demodulated phase containing the received PRBS chips is sent to a PRBS correlator at a fixed delay after a special packet announcing the transmission of a PRBS sequence. This packet allows us to reduce computational overhead by computing the correlation in a window near the estimated correlation peak time. The peak of the correlation signal is fit with a parabola to determine the timestamp to subsample precision. During calibration, we measured a 14 ps peak-to-peak bias on the subsample delay which arises from the parabolic fitting procedure. We apply an approximate correction using a low-order Fourier series of the measured periodic bias, such that any residual bias is well below the timing noise of the system.

After the master site timestamps are computed and are sent over the data link to the remote site, the remote site computes Eq. 1 to find the clock difference at a 2.2 kHz update rate. This difference is sent to a Kalman filter which provides robust estimates of $\Delta T_{AB}$ in the presence of signal dropouts [21]. The remote comb is then steered by sending the output of the Kalman filter to a simple PI loop filter on a direct digital synthesizer (DDS) mixed in with the comb stabilization lock to drive $\Delta T_{AB}$ to zero.

The bandwidth of this synchronization feedback is controlled by the dynamical response of the Kalman filter, whose optimal response time depends on the noise on the measured $\Delta T_{AB}$ versus the frequency random walk of clock's local oscillator. Clearly, the more stable the local oscillator, the lower the optimal synchronization feedback bandwidth. Effectively, by lowering the bandwidth, we average the measured time offset until its uncertainty is below the timing drift of the local oscillator. In our case, the local oscillator is a stable optical cavity, and the Kalman filter converges to a -3dB bandwidth of approximately 0.05 Hz to give optimal synchronized performance. If instead, the local oscillator was a quartz oscillator, then the Kalman filter settings would provide a somewhat higher synchronization bandwidth, as discussed later.

## 3. Results

Fig. 2a shows the time difference between the clocks during synchronization over 4 km distance, measured with out-of-loop timing data (see Figure 1a). The peak to peak wander over 8 hours is 16 ps with a standard deviation of 4 ps. As shown in Fig. 2b, to achieve this level of synchronization, the remote clock time is adjusted by ~ 1 µs. We next analyze these data in terms of the usual metrics of power-spectral density (PSD), and Allan deviations.

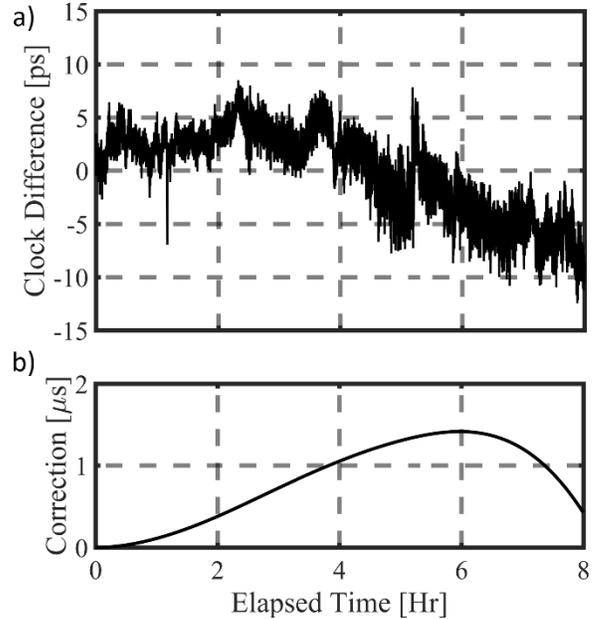

Fig. 2: a) Clock difference at 1 s sample rate and b) the clock correction applied to the remote clock under active synchronization over approximately 4 km of air. The standard deviation is 4 ps over 8 hours with a temperature-driven peak-to-peak wander of 16 ps. The system corrects for ~1 µs of timing excursion between the clocks over the measurement period, demonstrating a suppression of five orders of magnitude. Data is resampled to 1 Hz.

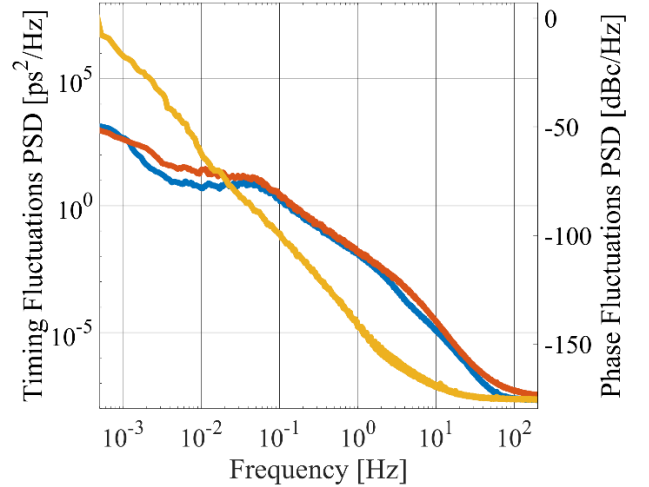

Fig. 3: Timing power spectral density of the clock time difference at 4 km (red) and over a shorted link (blue) under active synchronization and for a free-running system (yellow). Traces smoothed for visibility. The right axis shows the equivalent phase noise PSD for a nominal carrier frequency of 10 MHz.

The PSD for the relative timing noise between the clocks is shown in Fig. 3 for the data in Fig. 2. In addition, we show results for a shorted link under both free-running (non-synchronized) and synchronized operation. The free-running and synchronized traces cross at a location given by the Kalman filter steady-state response (lower than the synchronization bandwidth). Above the crossover, the oscillators (i.e. cavity stabilized lasers used for this measurement) have a timing PSD lower than the synchronization system. However, at lower Fourier frequencies below the crossover, the $f^{-4}$ relative noise of the oscillators dominates leading to a significant clock difference if not for the active synchronization system's feedback loop. Indeed, the free-running relative timing noise is suppressed by four orders of magnitude at 1 mHz by the synchronization. The integrated timing jitter from the synchronization bandwidth to 0.2 mHz is 1.5 ps for active synchronization over the 4-km path and 300 ps for the free-running case.

The fractional frequency uncertainty (e.g. MDEV) is shown for the same data in Fig. 4. At 1000 seconds, the synchronized clocks agree in frequency at the $10^{-15}$ level. The slightly higher results for the 4-km data over the shorted data is attributed to the larger temperature swings of 3 degrees Celsius seen by the transceivers due to the open windows.

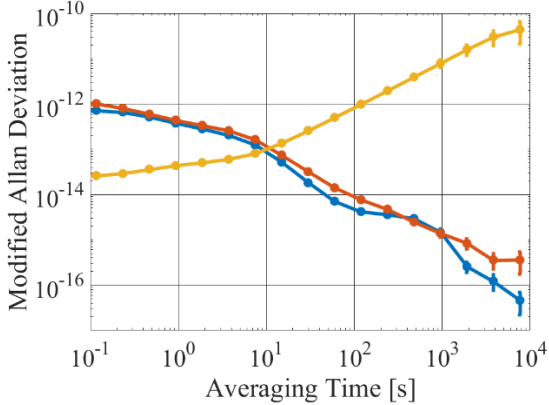

Fig. 4: Modified Allan Deviation at 4 km (red) and over a shorted link (blue) under active synchronization and for a free-running system (yellow). The inverse synchronization bandwidth of the system is where the traces cross at 10 s.

The synchronized clocks maintain a timing deviation, TDEV, (Fig. 5) below 1 ps beyond one hour. In our system, the local oscillators were cavity-stabilized lasers, which have low intrinsic phase noise. However, a practical system might instead use a simpler quartz oscillator. Fig. 5 also shows the expected free-running and synchronized performance for a remote clock based on a quartz oscillator, which reaches 1 ps for times from 10 s to 1 hr as well. This estimate is based on the measured performance of a quartz oscillator[16], an assumed $f^{-4}$ phase noise of the quartz oscillator, and optimally chosen Kalman filter parameters.

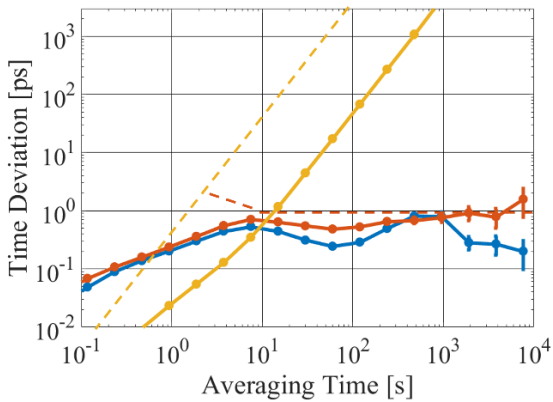

Fig. 5: Time deviation for synchronized clocks over 4-km link (red dots) and shorted link (blue dots), and time deviation for free-running unsynchronized clocks (yellow dots). The estimated free-running time deviation for remote clock based on a quartz oscillator is higher (dashed yellow line) but with synchronization should also drop to 1 ps (red dashed line) above 10 s.

## 4. Performance Limitations

In this section, we distinguish between the fast noise that sets the white noise floor on the measured clock time difference and the slow noise that causes time offset drifts over minutes-to-hours (e.g. see Fig. 2). The former depends on the signal-to-noise ratio of the coherently demodulated PRBS sequence, while the latter is due to out of loop thermal drifts and communications laser carrier frequency fluctuations (which we suppressed).

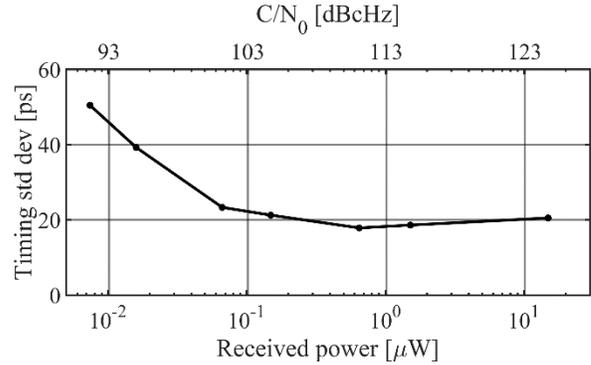

Fig 6: Standard deviation of the measured clock time offset vs. power (bottom axis) and carrier-to-noise ratio (top axis) at the 2 kHz update rate.

Fig. 6 shows the standard deviation of the measured time offset versus received power and carrier-to-noise ratio. A common oscillator was used for both sites in these measurements so we attribute all timing noise to the OTWTFT system. For these data, the standard deviation is calculated over a short time period and thus represents the fast noise. The level of this timing noise depends on the detection SNR of the coherently demodulated PRBS (as this sets the precision with which we can define the arrival of the PRBS sequence through correlation with the known pattern).

The detection SNR of the demodulated binary-phase shift keyed PRBS is limited by laser phase noise at high received powers and by additive white noise at low received powers. At powers above around 100 nW, the fast laser phase noise on the DFB lasers limits the detection SNR giving a timing noise floor of ~20 ps (over 10 seconds with no averaging) independent of received optical power. Below 100 nW, detector noise and shot noise limit the SNR so that the timing noise degrades with lower received optical power.

The location of this transition is due to the receive filter used, in this case designed to compromise between rejection of $f^{-2}$ laser phase noise and matched filtering of the received signal (which would give optimum performance in white phase noise only). Given sufficient detection power, a different choice of communication lasers or a higher data rate could reduce the timing noise floor further.

The primary driver of the slow timing drift is environmental. Any noise or drift in optical delays that are not common mode to bidirectional propagation will be incorrectly attributed to clock time offset. For example, the air temperature changed by 3 degrees Celsius over the measurement and this drove length changes in both the RF cabling and fiber optics. Under shorted link operation, the measured timing drift is correlated to the cycling of the room temperature from the AC unit.

In addition, any wander in the frequency offset between the communication lasers will cause timing noise due to the coupling of the carrier frequency and group delay. This systematic effect is a function of the group delay of the RF components in the send and receive chain. To minimize this effect, the two communication lasers are phase-locked based on the measured incoming communication signal itself to a frequency offset of 247 MHz by use of an acousto-optic modulator in addition to the slow temperature lock of the laser at the remote site. This frequency offset minimizes the timing noise as the combined group delays of the RF components is at a minimum. With the phase lock active, deviations from the minimum result in timing error below a few picoseconds, such that the group delay effect is well below the detection SNR and thermal limitations of the system.

## 5. Conclusion

We have shown a synchronization system that is compatible with an optical communications channel over a free space link. Traversing 4 km of air, this approach yields a residual time deviation below a picosecond from 10 seconds to 1 hour with a peak to peak wander of 16 picoseconds over 8 hours. The optical cavities used as local oscillators could be replaced with microwave oscillators with negligible degradation in performance.

This method provides a pragmatic trade-off between complexity and performance, leveraging the advantages of two-way free-space optical transfer to provide picosecond-level timing. In systems where picosecond-level synchronization is sufficient, this approach can provide a simpler, more compact, and distance-scalable alternative to comb-based OTWTFT. Furthermore, the picosecond timing precision is achieved in seconds, which enables real-time picosecond-level synchronization or time comparison across intermittent links.

If scaled up in power and aperture, the method should be applicable to future ground-to-space and intra-satellite time transfer [24,25]. Optical cross-links between future navigation satellites could lead to an improved global and navigation satellite system by enabling an actively synchronized clock network. For ground-to-satellite links, this approach could provide higher timing precision than the pulse position modulation approach used by NASA's Lunar Laser Communications Demonstration [19] at the cost of a modest increase in power.

Funding.
National Institute of Standards and Technology (NIST); Defense Sciences Office, DARPA (DSO, DARPA).

Acknowledgment.
We thank Martha Bodine and Mick Cermak for technical assistance.

## REFERENCES

1. M. A. Khalighi and M. Uysal, "Survey on Free Space Optical Communication: A Communication Theory Perspective," IEEE Commun. Surv. Tutor. **16**, 2231–2258 (2014).
2. F. Riehle, "Optical clock networks," Nat. Photonics **11**, 25–31 (2017).
3. J.-D. Deschênes, L. C. Sinclair, F. R. Giorgetta, W. C. Swann, E. Baumann, H. Bergeron, M. Cermak, I. Coddington, and N. R. Newbury, "Synchronization of Distant Optical Clocks at the Femtosecond Level," Phys. Rev. X **6**, 021016 (2016).
4. D. Kirchner, "Two-way time transfer via communication satellites," Proc. IEEE **79**, 983–990 (1991).
5. D. Kirchner, H. Ressler, P. Grudler, F. Baumont, C. Veillet, W. Lewandowski, W. Hanson, W. Klepczynski, and P. Uhrich, "Comparison of GPS Common-view and Two-way Satellite Time Transfer Over a Baseline of 800 km," Metrologia **30**, 183 (1993).
6. M. Fujieda, D. Piester, T. Gotoh, J. Becker, M. Aida, and A. Bauch, "Carrier-phase two-way satellite frequency transfer over a very long baseline," Metrologia **51**, 253 (2014).
7. D. Calonico, E. K. Bertacco, C. E. Calosso, C. Clivati, G. A. Costanzo, M. Frittelli, A. Godone, A. Mura, N. Poli, D. V. Sutyrin, G. Tino, M. E. Zucco, and F. Levi, "High-accuracy coherent optical frequency transfer over a doubled 642-km fiber link," Appl. Phys. B **117**, 979–986 (2014).
8. O. Lopez, F. Kéfélian, H. Jiang, A. Haboucha, A. Bercy, F. Stefani, B. Chanteau, A. Kanj, D. Rovera, J. Achkar, C. Chardonnet, P.-E. Pottie, A. Amy-Klein, and G. Santarelli, "Frequency and time transfer for metrology and beyond using telecommunication network fibres," Comptes Rendus Phys. **16**, 531–539 (2015).
9. Ł. Śliwczyński, P. Krehlik, A. Czubla, Ł. Buczek, and M. Lipiński, "Dissemination of time and RF frequency via a stabilized fibre optic link over a distance of 420 km," Metrologia **50**, 133 (2013).
10. J. Kim, J. A. Cox, J. Chen, and F. X. Kärtner, "Drift-free femtosecond timing synchronization of remote optical and microwave sources," Nat. Photon **2**, 733–736 (2008).
11. S. Droste, F. Ozimek, T. Udem, K. Predehl, T. W. Hänsch, H. Schnatz, G. Grosche, and R. Holzwarth, "Optical-Frequency Transfer over a Single-Span 1840km Fiber Link," Phys. Rev. Lett. **111**, 110801 (2013).
12. P. Krehlik, H. Schnatz, and Ł. Śliwczyński, "A Hybrid Solution for Simultaneous Transfer of Ultrastable Optical Frequency, RF Frequency, and UTC Time-Tags Over Optical Fiber," IEEE Trans. Ultrason. Ferroelectr. Freq. Control **64**, 1884–1890 (2017).
13. L. C. Sinclair, W. C. Swann, H. Bergeron, E. Baumann, M. Cermak, I. Coddington, J.-D. Deschênes, F. R. Giorgetta, J. C. Juarez, I. Khader, K. G. Petrillo, K. T. Souza, M. L. Dennis, and N. R. Newbury, "Synchronization of clocks through 12 km of strongly turbulent air over a city," Appl. Phys. Lett. **109**, 151104 (2016).
14. S. Chen, F. Sun, Q. Bai, D. Chen, Q. Chen, and D. Hou, "Sub-picosecond timing fluctuation suppression in laser-based atmospheric transfer of microwave signal using electronic phase compensation," Opt. Commun. **401**, 18–22 (2017).
15. J. Anderson, N. Barnwell, M. Carrasquilla, J. Chavez, O. Formoso, A. Nelson, T. Noel, S. Nydam, J. Pease, F. Pistella, T. Ritz, S. Roberts, P. Serra, E. Waxman, J. W. Conklin, W. Attai, J. Hanson, A. N. Nguyen, K. Oyadomari, C. Priscal, J. Stupl, J. Wolf, and B. Jaroux, "Sub-nanosecond ground-to-space clock synchronization for nanosatellites using pulsed optical links," Adv. Space Res. (2017).
16. E. Samain, P. Exertier, P. Guillemot, P. Laurent, F. Pierron, D. Rovera, J. Torre, M. Abgrall, J. Achkar, D. Albanese, C. Courde, K. Djeroud, M. L. Bourez, S. Leon, H. Mariey, G. Martinot-Lagarde, J. L. Oneto, J. Paris, M. Pierron, and H. Viot, "Time Transfer by Laser Link - T2L2: Current status and future experiments," in *Frequency Control and the European Frequency and Time Forum (FCS), 2011 Joint Conference of the IEEE International* (2011), pp. 1–6.
17. K. U. Schreiber, I. Prochazka, P. Lauber, U. Hugentobler, W. Schafer, L. Cacciapuoti, and R. Nasca, "Ground-based demonstration of the European Laser Timing (ELT) experiment," IEEE Trans. Ultrason. Ferroelectr. Freq. Control **57**, 728–737 (2010).
18. O. Lopez, A. Kanj, P.-E. Pottie, D. Rovera, J. Achkar, C. Chardonnet, A. Amy-Klein, and G. Santarelli, "Simultaneous remote transfer of accurate timing and optical frequency over a public fiber network," Appl. Phys. B **110**, 3–6 (2013).
19. D. V. Murphy, J. E. Kansky, M. E. Grein, R. T. Schulein, M. M. Willis, and R. E. Lafon, "LLCD operations using the Lunar Lasercom Ground Terminal," in *Free-Space Laser Communication and Atmospheric Propagation XXVI* (International Society for Optics and Photonics, 2014), Vol. 8971, p. 89710V.
20. A. Biswas, M. Srinivasan, S. Piazzolla, and D. Hoppe, "Deep space optical communications," in *Free-Space Laser Communication and Atmospheric Propagation XXX* (International Society for Optics and Photonics, 2018), Vol. 10524, p. 105240U.
21. H. Bergeron, L. C. Sinclair, W. C. Swann, C. W. Nelson, J.-D. Deschênes, E. Baumann, F. R. Giorgetta, I. Coddington, and N. R. Newbury, "Tight real-time synchronization of a microwave clock to an optical clock across a turbulent air path," Optica **3**, 441 (2016).
22. M. Cvijetic and I. Djordjevic, *Advanced Optical Communication Systems and Networks* (Artech House, 2013).
23. J. H. Shapiro, "Reciprocity of the Turbulent Atmosphere," J Opt Soc Am **61**, 492–495 (1971).
24. H. Kaushal and G. Kaddoum, "Optical Communication in Space: Challenges and Mitigation Techniques," IEEE Commun. Surv. Tutor. **19**, 57–96 (2017).
25. P. Berceau, M. Taylor, J. M. Kahn, and L. Hollberg, "Space-Time Reference with an Optical Link," Class. Quantum Gravity **33**, 135007 (2016).